# MRPC3b mass production for CBM-TOF and eTOF at STAR

**D. Hu,**[abc] **D. Sauter,**[c] **Y. Sun,**[ab,*] **I. Deppner,**[c,†] **N. Herrmann,**[c,‡] **J. Brandt,**[c] **H. Chen,**[ab] **E. Lavrik,**[d] **C. Li,**[ab] **M. Shao,**[ab] **C. Simon,**[c] **Z. Tang,**[ab] **X. Wang,**[ab] **Ph. Weidenkaff,**[c] **Y. Zhang,**[ab] **J. Zhou,**[ab] **and H. Zeng**[ab]

[a] *Department of Modern Physics, University of Science and Technology of China (USTC), 96 Jinzhai Road, Hefei 230026, China*

[b] *State Key Laboratory of Particle Detection and Electronics, USTC-IHEP, 96 Jinzhai Road, Hefei 230026, China*

[c] *Physikalisches Institut der Universität Heidelberg, Im Neuenheimer Feld 226, Heidelberg, Germany*

[d] *Physikalisches Institut der Universität Tuebingen, Auf der Morgenstelle 14, Tuebingen Germany*

*E-mail*: sunday@ustc.edu.cn
*E-mail*: deppner@physi.uni-heidelberg.de
*E-mail*: herrmann@physi.uni-heidelberg.de

ABSTRACT: The Compressed Baryonic Matter (CBM) spectrometer aims to study strongly interacting matter under extreme conditions. The key element providing hadron identification at incident energies between 2 and 11AGeV in heavy-ion collisions at the SIS100 accelerator is a Time-of-Flight (TOF) wall covering the polar angular range from 2.5°–25° and full azimuth. CBM is expected to be operational in the year 2024 at the Facility for Anti-proton and Ion Research (FAIR) in Darmstadt, Germany. The existing conceptual design foresees a 120 m$^2$ TOF-wall composed of Multi-gap Resistive Plate Chambers (MRPC) which is subdivided into a high rate region, a middle rate region and a low rate region. The MRPC3b Multistrip-MRPCs, foreseen to be integrated in the low rate region, have to cope with charged particle fluxes up to 1 kHz/cm$^2$ and therefore will be constructed with thin float glass (0.28 mm thickness) as resistive electrode material. In the scope of the FAIR phase 0 program it is planned to install about 36% of the MRPC3b counters in the east endcap region of the STAR experiment at BNL as an upgrade for the Beam Energy Scan campaign (BESII) in 2019/2020.

---

[*] Corresponding author.
[†] Corresponding author.
[‡] Corresponding author.

# Contents

## 1. Introduction

A large-area time-of-flight (TOF) wall based on high-resolution MRPC technology has been designed [1] for the Compressed Baryonic Matter (CBM) experiment [2] at the Facility for Antiproton and Ion Research (FAIR) [3], planned at the current GSI site near Darmstadt, Germany. The key element providing hadron identification at incident energies between 2 and 11AGeV at SIS100 is a Time-of-Flight (TOF) wall covering the polar angular range from 2.5°– 25° and full azimuth. The existing conceptual design foresees a 120 $m^2$ TOF-wall composed of Multi-gap Resistive Plate Chambers (MRPC) [4] which is subdivided into a high rate region, a middle rate region and a low rate region. In addition to the requested rate capability all MRPC units have to fulfill challenging requirements concerning time resolution and efficiency in order to allow for a good performance of their main task in the experiment, the separation of pion kaon and proton in the momentum range of up to 4 GeV/c at 3 sigma separation. The TOF system's overall time resolution has to be better than 80ps including electronics/digitization jitter and the detection efficiency has to be better than 95%. These requirements have to be met at the anticipated charged-particle flux that drops from a few tens of kHz/ $cm^2$ close to the beam pipe — depending on the incident beam energy — to below 1kHz/ $cm^2$ in the wall's periphery. Actually, for the outer region of about 50% of the total TOF-wall active area, rate capabilities up to 1 kHz/ $cm^2$ are sufficient for the anticipated high rate running scenario with an Au + Au interaction rate of $10^7$ $s^{-1}$ at 11 AGeV. Counters with this moderate rate capability will be positioned in the blue marked region of the current conceptual design shown in Fig. 1. To be cost effective the peripheral MRPC modules will be equipped with float glass MRPCs (MRPC3b and MRPC4) instead of using the more expensive low-resistive electrode material [5]. Prototypes of MRPC3b have been build and were used in a proof-of-principle study [6] in two test beam campaigns at the SPS accelerator at CERN.

As a next step in the preparation of the CBM-TOF wall that has to be operational in 2024 at FAIR, the CBM modules will be used in the STAR experiment in 2019 and 2020 as part of the so called FAIR phase 0 program. The CBM collaboration institutions: Heidelberg, Darmstadt,

Tsinghua, CCNU, USTC and the STAR collaboration are cooperating in installing, commissioning, and operating a "wheel" of CBM TOF detectors mounted on the inside face of the STAR east side pole tip. The conceptual design of the setup is shown in Fig.2. This endcap TOF (eTOF) detector is extending STAR's particle identification (PID) in the intermediate momentum range up to at least a pseudorapidity η of -1.5 [7]. The installation of eTOF is accompanied by two more upgrades of the STAR detector targeted towards STAR's beam energy scan phase II (BES-II) physics runs: the inner part of the TPC is upgraded with additional readout channels (iTPC upgrade) [8] and the Event Plane Detector (EPD) is added to the experiment [9]. Combined with the iTPC upgrade, the eTOF upgrade will provide additional particle identification capability in the forward direction at STAR, which will especially enhance the Fixed Target program at BES-II.

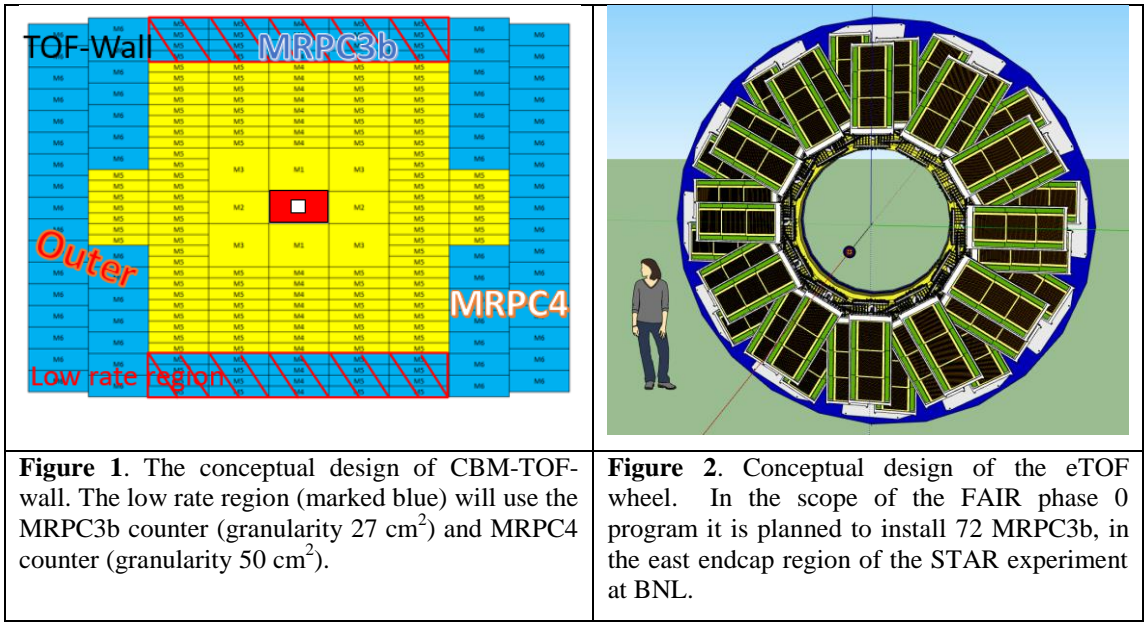

**Figure 1**. The conceptual design of CBM-TOF-wall. The low rate region (marked blue) will use the MRPC3b counter (granularity 27 $cm^2$) and MRPC4 counter (granularity 50 $cm^2$).

**Figure 2**. Conceptual design of the eTOF wheel. In the scope of the FAIR phase 0 program it is planned to install 72 MRPC3b, in the east endcap region of the STAR experiment at BNL.

In this proceeding we present the design of the MRPC3b counter and report on the CBM-TOF MRPC3b mass production status at USTC, as well as on the quality control and assurance (QC&QA) procedure. The MRPC3b counters are integrated into modules at Heidelberg University and all necessary counter information is stored in a component database. The structure of this database is explained in this contribution.

## 2. The MRPC3b counter

The MRPC3b counter is a multi-gap RPC with 32 read out strips which are read out on both sides. Figure 3 shows the strips layout of the MRPC3b counter. The strip width is 7mm with a spacing of 3 mm and a length of 27.6mm. The strip width to pitch ratio is carefully designed in order to avoid impedance mismatches. The MRPC3b counter has in total 10 gas gaps arranged in a double-stack configuration mirrored with respect to the central electrode. Each gas gap is 0.230 mm in size, defined by nylon fishing lines as shown in Figure 4 placed between 6 float glass plates each 0.28 mm thick. The glass plates have the dimension of 33 x 28.6 x 0.028 cm$^3$ while the active area is 32 x 27.6 cm$^2$.

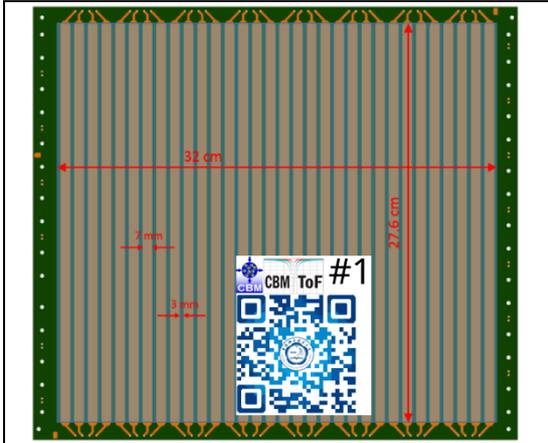 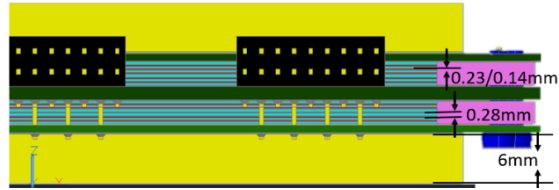

**Figure 3**. MRPC3b readout electrode and superimposed a QR code sticker. The active area is 32 x 27.6 cm$^2$. During the mass production information is stored in a component data base which can be accessed by the QR code.

**Figure 4**. This drawing shows the side elevation of MRPC3b. The gap size is 0.230 mm while the thickness of the float glass plate is 0.28 mm. The structural support of the MRPC is ensured by a 6 mm thick honeycomb plate.

## 3. Mass production and QC&QA procedure

During the last year we started at the TOF laboratory in USTC the preparation of the mass production for the MRPC3b counters. In order to guarantee the quality of the detectors the class of the clean room got improved. Currently we have a clean room (conf. Figure 5) of 200 m$^2$ size with a cleanness level of 100 k. The mounting table has additional air drains which improves the cleanness level even more. The temperature and humidity of the room is kept stable at 22 $\pm$2 ℃ and ≤40% respectively.

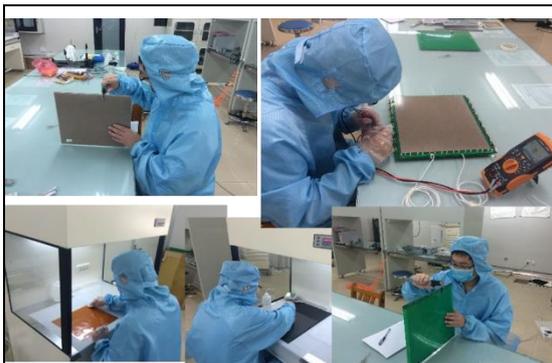 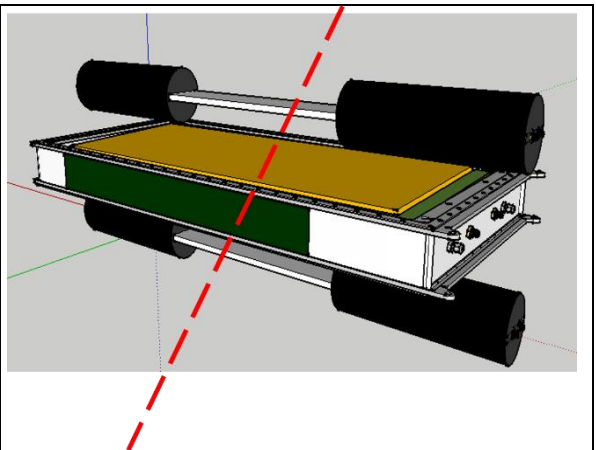

**Figure 5**. Clean room and clean desk for MRPC mass production at the USTC site.

**Figure 6**. New cosmic test stand

A cosmic test stand system is built for quality control in the laboratory of the USTC, as shown in Fig.6. For the mass production, two systems of the cosmic stand are built with different readout systems. One system is developed by USTC electronic group, while another system uses the TRB system [11] which was developed by GSI, Germany. Besides the test system, the QA procedure is equally important. A diagram of the QA procedure at various production stage is shown in Fig. 7. The quality of all components and manufacturing steps are constantly monitored and stored in a MRPC3b production data base. This includes step-by-step MRPC production manual, flow diagram of MRPC quality control, manufacture cards, check cards, manufacture log sheets and record tables for test results. A QR-code glued on the detector surface is generated containing a link with the content of the MRPC3b production data base.

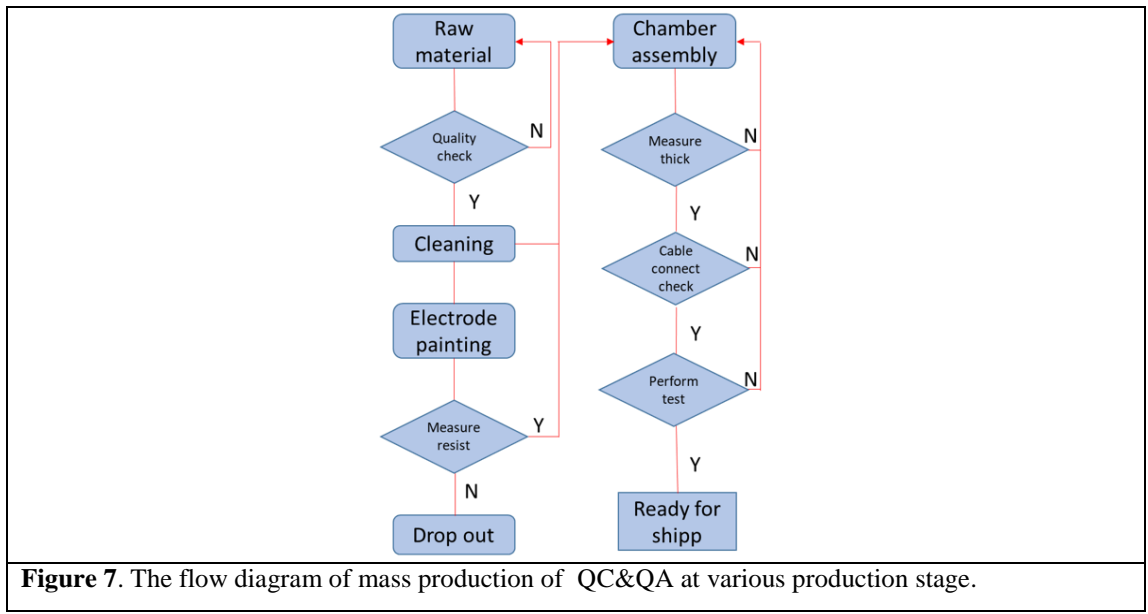

**Figure 7**. The flow diagram of mass production of QC&QA at various production stage.

By now 10 out of 80 MRPC3b counters have been produced and the mass production will continue until end of July, 2018. These 10 counters are tested at TOF laboratory in Heidelberg University. This pre-series will be installed in modules which are foreseen for the eTOF wheel at STAR.

In order to find the optimum operation voltage of the MRPC, a wide-range HV scan was done and the efficiency was checked for modified eTOF module described in section 4 at Heidelberg University. 2 MRPC counters are placed on top of each other in each module. The cosmic test stand is shown in Fig. 8a and set up to test 6 MRPC3 counters at the same time. The entire system runs in free streaming data acquisition mode. The data are analyzed by a simple track - reconstruction method. Hits in the top module and the bottom module are used as reference and the connecting straight line defines a track candidate. Hits in the intermediate chambers that are found in the vicinity of the track candidate are associated to the track. Up to 6 hits can be collected for a track. Comparing the time of a given hit to the average expectation of the other 5 hits allows to extract a system time resolution. The result is shown in Fig. 8b together with the counter efficiency. The latter is derived from the comparison of 6-hit to 5-hit tracks, i.e. whenever a track with 5 hits intersect the active area of a counter in one of the intermediate planes, a hit in this counter is expected. The ratio of events finding a hit at this spot to the total number of 5- and 6-hit tracks defines the efficiency of the counter.

Preliminarily the detection efficiency of eTOF –MRPC is above 95% when the applied HV is higher than $\pm 6$kV and the counter time resolution is then better than 60 ps (including the electronic jitter).

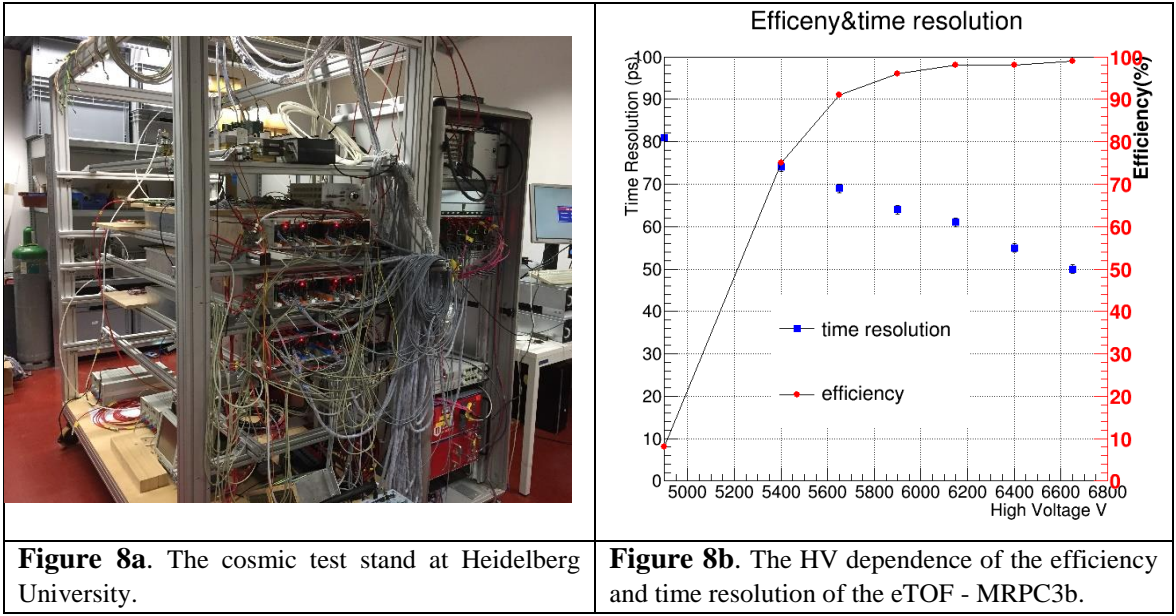

**Figure 8a**. The cosmic test stand at Heidelberg University.

**Figure 8b**. The HV dependence of the efficiency and time resolution of the eTOF - MRPC3b.

## 4. eTOF module – integration

Within the FAIR Phase 0 program, 10% of the full CBM TOFs sytem including read-out chain will be install as eTOF subsystem at STAR/RHIC (BES II 2019/2020). The entire eTOF system comprises 36 modules, 3 layers, 12 sectors, 6912 channels, and the corresponding readout electronics. Figure 9 shows the eTOF module and the eTOF module's electronic. Each module's active area is 97 x 27 cm$^2$ subdivided into 96 readout strips. Each module has 192 readout channels. The eTOF module houses front-end electronics boards (FEE) that implement the PADIX discriminator for MRPC signal processing. PADI is a general purpose PreAmplifier-Discriminator ASCI to be used as Front-End-Electronics (FEE) for readout of timing Resistive Plates Chamber (RPC) detectors in the CBM experiment at FAIR, designed by CBM-TOF group [12]. The signals of the FEE PADI boards located in the modules are fed into the GET4 - TDC that is attached to the module. The GSI Event-Drive TDC GET4 is a high resolution low power event-driven TDC for the CBM-Time of Flight detector readout [13]. For testing purposes, one module equipped with MRPC3b counters (see Fig. 10) was installed at STAR in January 2018 and operated during the run2018 (Mar. –Jun.).

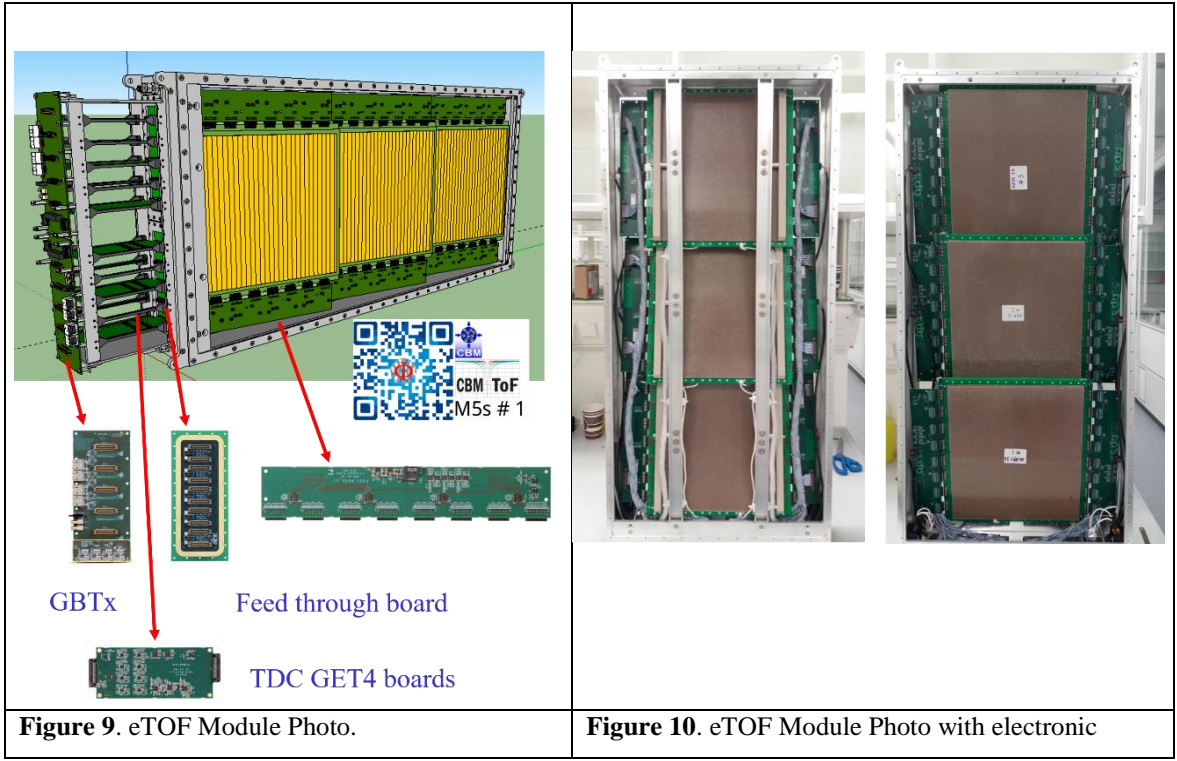

| Figure 9. eTOF Module Photo. | Figure 10. eTOF Module Photo with electronic |

## 5. Component database for CBM-TOF

The CBM-TOF component database is based on software packages i.e. Sqlite3 [14], Fairroot [15], FairDB [16], nodejs, html and jsroot. It will be used to store the basic information accumulated during mass production i.e. module assembly information, electronic information, channel map and counters information. The following points were considered important for the design of the database. The component database should have a GUI to enter data, should be easy to operate, should be conveniently modifiable, should be able to store images and histograms, should display histograms in the browser, and should be based on CBMROOT [17]. The structure of the CBM-TOF database is developed based on the FairDB kernel. Using a webpage/browser interface to FairDB input data from the webpage can be directly stored in the sqlite3 database. The CBM-TOF database also provides the feature that the database code can be directly generated through the webpage. The storage and retrieval of histograms and images is also successfully implemented as shown in Fig. 11.

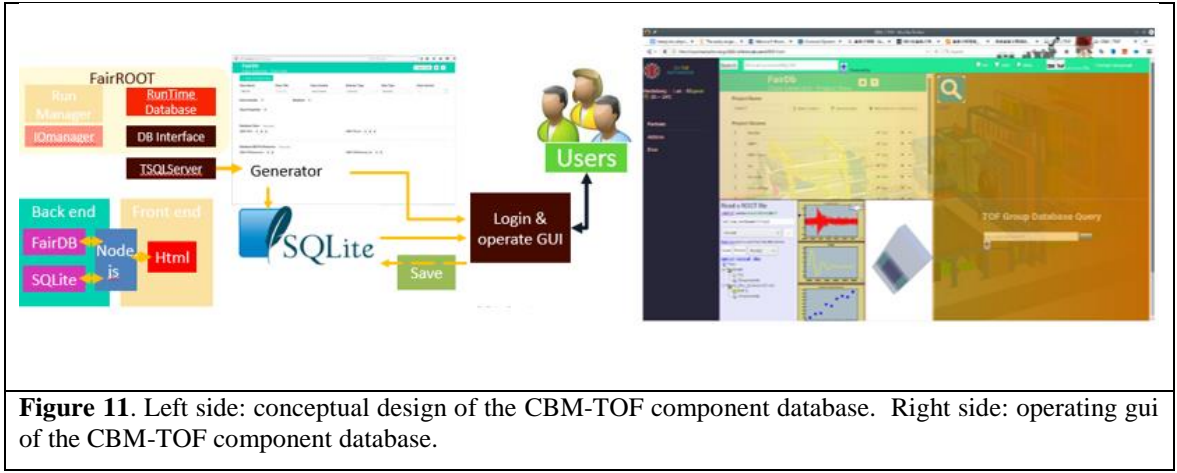

**Figure 11**. Left side: conceptual design of the CBM-TOF component database. Right side: operating gui of the CBM-TOF component database.

## 6. Summary and outlook

The CBM Time-of-Flight system is developed by 8 institutions from China, Germany Romania and Russia. At USTC the MRPC3b detector using float glass electrodes is developed which is foreseen for the low rate region of the CBM TOF wall. In the context of the CBM FAIR phase 0 program modules housing the MRPC3b counter will be installed and operated in the STAR experiment at BNL during the beam energy scan II program.

## Acknowledgments

This work thank the group of electronic of USTC for their support. This project is supported by National Natural Science Foundation of China (U1232206), International Cooperation and Exchanges Project of NSFC (11420101004) and National Program on Key Basic Research Project of China-973 Program (2015CB856902) and the German BMBF project 05P15VHFC1.